# Thermal Properties of Isotopically Engineered Graphene

Shanshan Chen[1,2], Qingzhi Wu[2], Columbia Mishra[2], Junyong Kang[1], Hengji Zhang[3], Kyeongjae Cho[3,4], Weiwei Cai[1, 2*], Alexander A. Balandin[5*] and Rodney S. Ruoff[2*]

[1]Department of Physics, Fujian Key Laboratory of Semiconductor Materials and Application, Xiamen University, Xiamen 361005 China

[2]Department of Mechanical Engineering and the Materials Science and Engineering Program, University of Texas at Austin, Austin, TX 78712 USA

[3]Department of Materials Science and Engineering and Department of Physics, University of Texas at Dallas, Richardson, TX 75080 USA

[4]School of Mechanical and Aerospace Engineering, Seoul National University, Seoul 151-742 Republic of Korea

[5]Department of Electrical Engineering and Materials Science and Engineering Program, University of California at Riverside, Riverside, CA 92521 USA

[*]To whom correspondence should be addressed.
E-mail: wwcai@xmu.edu.cn, balandin@ee.ucr.edu, r.ruoff@mail.utexas.edu

**In addition to its exotic electronic properties [1-2] graphene exhibits unusually high intrinsic thermal conductivity [3-6]. The physics of phonons – the main heat carriers in graphene – was shown to be substantially different in two-dimensional (2D) crystals, such as graphene, than in three-dimensional (3D) graphite [7-10]. Here, we report the first experimental study of the isotope effects on the thermal properties of graphene. Isotopically modified graphene containing various percentages of $^{13}C$ were synthesized by chemical vapor deposition (CVD). The regions of different isotopic composition were parts of the same graphene sheet to ensure uniformity in material parameters. The thermal conductivity, $K$, of isotopically pure $^{12}C$ (0.01% $^{13}C$) graphene determined by the *optothermal Raman* technique [3-7, 10], was higher than 4000 W/mK at the measured temperature $T_m$~320 K, and more than a factor of two higher than the value of $K$ in a graphene sheets composed of a 50%-50% mixture of $^{12}C$ and $^{13}C$. The experimental data agree well with our molecular dynamics (MD) simulations, corrected for the long-wavelength phonon contributions via the Klemens model. The experimental results are expected to stimulate further studies aimed at better understanding of thermal phenomena in 2D crystals.**

Naturally occurring carbon materials are made up of two stable isotopes of $^{12}C$ (abundance ~99%) and $^{13}C$ (~1%). The change in isotope composition modifies dynamic properties of crystal lattices and affects their thermal conductivity, $K$. The isotopically purified materials are characterized by enhanced $K$ [11-13]. The knowledge of isotope effects on $K$ is valuable for understanding the phonon transport. The isotope composition affects directly the phonon relaxation via the phonon mass-difference scattering. The phonon-scattering rate on point defects, $1/\tau_P$, is given as [14-16] $1/\tau_P \propto V_0(\omega^\alpha/\upsilon^\beta)\Gamma$, where $V_0$ is the volume per one atom in the crystal lattice, $\omega$ is the phonon frequency, $\upsilon$ is the phonon group velocity, $\Gamma$ is the strength of the phonon - point defect scattering, $\alpha$=3 (4) and $\beta$=2 (3) for 2D (3D) system, correspondingly. In the perturbation theory $\Gamma$ is written as [14-16]

$$\Gamma = \sum_i f_i \left[ (1 - M_i/\overline{M})^2 + \varepsilon(\gamma(1 - R_i/\overline{R}))^2 \right] \quad (1)$$

Here $f_i$ is the fractional concentration of the substitutional foreign atoms, e.g. impurity, defect or isotope atoms, $M_i$ is the mass of the $i$th substitutional atom, $\overline{M} = \sum_i f_i M_i$ is the average atomic mass, $R_i$ is the Pauling ionic radius of the $i$th foreign atom, $\overline{R} = \sum_i f_i R_i$ is the average radius, $\gamma$ is the Gruneisen parameter that characterizes the anharmonicity of the lattice, and $\varepsilon$ is a phenomenological parameter. The mass of a foreign atom (impurity, vacancy, defect or isotope) is well known while the local displacement $\Delta R = \overline{R} - R_i$ due to the atom radius or bond-length difference is usually not known.

From the kinetic transport theory, the thermal conductivity of graphene can be presented as $K = (1/2)Cv\Lambda$, where $C$ is the specific heat capacity and $\Lambda$ is the phonon mean free path (MFP). Assuming that $\Lambda$ is limited by the anharmonic phonon Umklapp and point-defect scattering it can be written as $\Lambda^{-1} = \Lambda_U^{-1} + \Lambda_P^{-1}$, where $\Lambda = v\tau_U$ is the Umklapp-limited MFP, $\Lambda = v\tau_P$ is the point-defect limited MFP, $\tau_U$ is the Umklapp-limited phonon life-time, $\tau_P$ is the point-defect-limited phonon life-time, which can be affected by isotopes, defects, impurities or vacancies. From these considerations and Eq. (1), one can see that the phonon-isotope scattering is unique in a sense that unlike impurity or defect scattering it involves only the well-defined mass-difference term, $\Delta M = \overline{M} - M_i$, without the ambiguous volume or bond-strength difference term, $\Delta R = \overline{R} - R_i$ and $\varepsilon$.

As the system dimensionality changes from 3D to 2D, $1/\tau_P$ undergoes additional modification owing to the different phonon density of states (DOS). The change in the phonon DOS reveals itself via dependence of $1/\tau_P$ on $\omega$ and $v$. Thus, the isotope effects in graphene are particularly important for understanding its thermal properties and, more generally, for development of theory of heat transport in low-dimensional systems. Until today, experimental studies of isotope effects in graphene were not possible because of unavailability of the proper samples. Theoretical studies of the isotopically enriched graphene just started [8, 17] and experimental data is needed for model validation.

The samples used in this study were large-area high-quality monolayer graphene grown by CVD on the interior surfaces of Cu foil enclosures [18]. The characteristic grain size in our samples was determined to be over ~200 μm. In order to avoid potential sample-to-sample variation, we used the isotope-labeling technique [18-19] to synthesize single-layer graphene with the regions of 0.01%, 1.1%, 50%, and 99.2% of $^{13}$C by turning on and off, sequentially, 99.99% $^{12}$CH$_4$, the 50:50 $^{12}$CH$_4$/$^{13}$CH$_4$ mixture, normal CH$_4$, and 99.2 % $^{13}$CH$_4$ gases. The isotope content in the localized graphene regions mirrored the dosing sequence employed. The isotope modified regions were verified with the micro-Raman spectroscopy using the distinctive phonon mode signatures of the isotope-engineered materials [18-20]. The graphene samples were transferred to a Au-coated silicon nitride (SiN$_x$) membrane with pre-fabricated 100×100 array of 2.8-μm diameter holes. A part of the same graphene film was transferred to a Si wafer with 285-nm-thick SiO$_2$ layer [21]. Details of the isotope-engineered graphene preparation can be found in the *Methods* section.

Raman spectroscopy of graphene transferred to the 285-nm SiO$_2$/Si wafer was performed under 532-nm laser light excitation (Figure 1). The distinctive Raman peaks for different $^{12}$C/$^{13}$C ratios allow for an accurate mapping of the isotopically labeled graphene regions [18-19] [1,2] [1,2]. The Raman 2D-band position map over a 120×120 μm$^2$ area reveals the "six-lobe" shape with the growing edges that resemble dendrites consistent with our previous observations [18]. Figure 1a is the Raman peak position mapping (2530~2730 cm$^{-1}$) of four regions that have different $^{13}$C-isotope content. The color bar at the bottom indicates the dose sequence. The G-peak and 2D-band positions in Raman spectra of graphene with 0.01%, 50%, 1.1%, and 99.2% $^{13}$C-isotope are presented in Figure 1b. They are recorded from different sample spots marked by the corresponding colored circles in Figure 1a. The Brillouin-zone-center optical-phonon frequency $\omega$ varies with the atomic mass, $M$, as $\omega \propto M^{-1/2}$ making the Raman shift for $^{13}$C approximately $(12/13)^{-1/2}$ times smaller than that for $^{12}$C [22-24]. The latter explains the observed sequence of the Raman peak positions in

Figure 1b. The experimental difference between the lowest 99.2% $^{13}$C peak and the highest 0.01% $^{13}$C peak is ~64 cm$^{-1}$, which is in agreement with the theory, and attests for the high quality of our isotopically modified graphene.

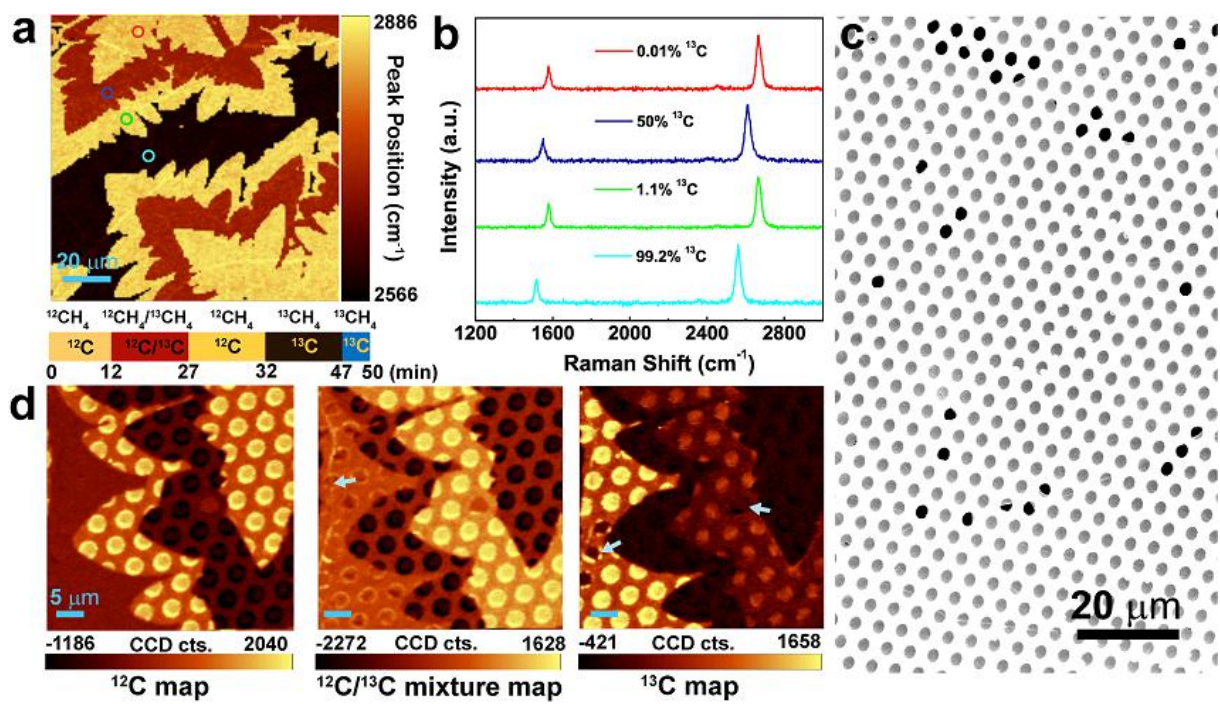

**Figure 1:** Micro-Raman characteristics of the isotopically modified graphene. (a) Raman map (2530~2730 cm$^{-1}$) of the 2D bands. The color bar shown in the bottom of (a) indicates the dosing sequence for the four regions in (a). (b) Raman spectra of graphene measured at the positions labeled by the colored circles in (a). (c) SEM image and (d) Raman maps of graphene transferred onto the SiN$_x$ holey membrane. (d) From left to right, three Raman maps showing the integrated intensity of the 2D band of 0.01 % $^{13}$C (2630~2730 cm$^{-1}$), the 50:50 $^{12}$C/$^{13}$C mixture (2560~2660 cm$^{-1}$) and 99.2% $^{13}$C (2530~2630 cm$^{-1}$) graphene, respectively. Arrows in (d) indicate wrinkles and cracks in some samples, which were excluded from the thermal measurements.

A different portion of the same $^{13}$C-labeled monolayer graphene sheet was transferred onto a Au-coated SiN$_x$ holey membrane. Figure 1c shows a scanning electron microscopy (SEM, FEI Quanta-600) image of graphene on the SiN$_x$ holey membrane. For a few holes in the array, the graphene film is broken or wrinkled. Before the thermal measurements, Raman mapping with a laser power of ~8 mW (on the sample surface) was carried out on the selected graphene areas, which could heat the graphene membranes up to about 600 K. The procedure was implemented both to further remove the PMMA residue and to verify the $^{13}$C concentration in the suspended graphene membranes. All portions of the sample were subjected to the same treatment.

Figure 1d shows 40×40 μm$^2$ integrated intensity Raman maps of the 2D-band of the $^{12}$C (2630~2730 cm$^{-1}$), the 50:50 $^{12}$C/$^{13}$C mixture (2560~2660 cm$^{-1}$), and of the $^{13}$C (2530~2630 cm$^{-1}$), respectively. The bright region in the $^{12}$C-map from right-to-left corresponds to 0.01% $^{13}$C graphene grown at the beginning and 1.1% $^{13}$C grown at the third dosing sequence. The bright region in the $^{12}$C/$^{13}$C-mixture map and $^{13}$C-map corresponds to 50% $^{13}$C and 99.2% $^{13}$C graphene grown in the second and last dosing sequence, respectively. The Raman mapping was used to avoid the sample parts with wrinkles and cracks [18, 19]. The arrow in the $^{12}$C/$^{13}$C-mixture map shows a wrinkle in the suspended graphene film while the arrows in the $^{13}$C map indicate places where graphene either does not cover or only partially covers the holes in the SiN$_x$ membrane (Figure 1d).

The thermal conductivity $K$ of the suspended isotopically modified graphene layers was measured by the non-contact optothermal Raman technique [3-7, 10, 25]. Typical results for $K$ as a function of the measured temperature $T_m$ (extracted from the Raman data) for four regions with different isotope composition are plotted in Figure 2. The isotopically-pure graphene ($^{13}$C 00.1%) has the highest $K$ of ~4120±1410 W/mK at $T_m$~320 K. For comparison, the layer of the natural abundance graphene ($^{13}$C 1.1%) of the same size reveals $K$≈2600±658 W/mK at $T_m$~330 K. The near room-temperature $K$ for the natural abundance graphene is consistent with previous studies within the experimental uncertainty [3-7, 10]. Graphene with $^{13}$C-content increased to 99.2% has a similar $K$ as the natural abundance graphene. The latter is explained by the fact that in both cases the amount of "isotope impurities", which give rise to the phonon mass-difference scattering (see Eq. (1)), are about the same ~1% of the total carbon atoms. The extrapolated $K$ values at 300 K are 4419 W/mK, 2792 W/mK, 2197 W/mK and 2816 W/mK for 0.01% $^{13}$C ("isotopically pure $^{12}$C"), 1.1% $^{13}$C (natural graphene), 50% $^{13}$C ("isotopically disordered" graphene), and 99.2% $^{13}$C ($^{13}$C enriched graphene), respectively. Comparing at the same temperature, $K$ is ~58% enhanced in the isotopically pure $^{12}$C graphene with respect to the natural graphene near room temperature.

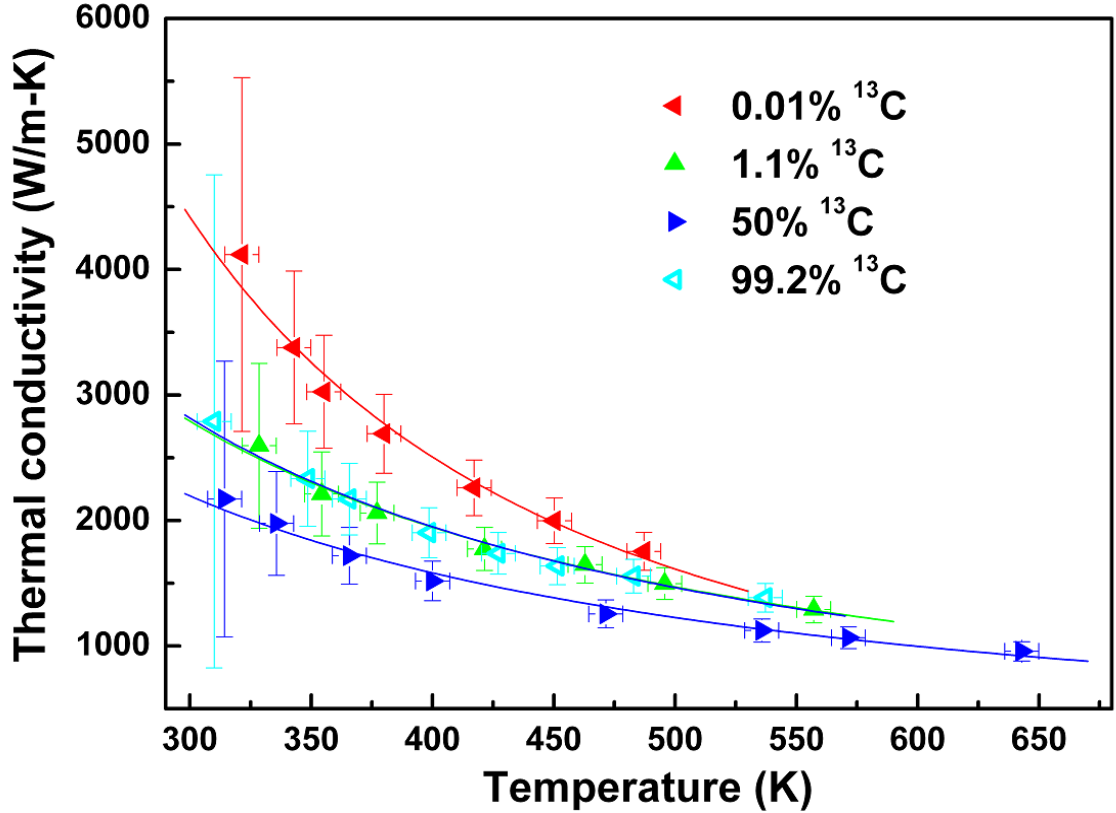

**Figure 2:** Thermal conductivity $K$ of the suspended graphene film with $^{13}$C isotope concentrations of 0.01%, 1.1% (natural abundance), 50% and 99.2%, respectively, as a function of the temperature measured with the micro-Raman spectrometer. The solid lines are a guide to the eye only. The experimental errors were estimated via the square-root-sum error propagation approach including the following error sources: the Raman peak position calibration, temperature resolution of the Raman measurement method, and the uncertainty of the laser absorption.

One can see from Figure 2 that the lowest $K$ is observed in graphene characterized by the strongest mixture of the isotopes ($^{13}$C 50%), which is expected from Eq. (1). The evolution of $K$ with the isotope content is mostly produced by changes in the phonon – point defect scattering rate $1/\tau_P$ via the mass-difference term $\Delta M = \overline{M} - M_i$. The phonon $\upsilon$ and mass density do not undergo substantial modification with the isotope composition. The relative change in the phonon velocity $\upsilon_{^{12}C}/\upsilon_{natural}$ is related to the mass densities of the respective lattices $\upsilon_{^{12}C}/\upsilon_{natural} = (M_{natural}/M_{^{12}C})^{1/2}$. Removal of 1% $^{13}$C in natural diamond causes the velocity to increase only by a factor of 1.0004 [13], which cannot account for the observed ~58% change in *K*.

To verify the reproducibility and minimize the data uncertainty, we prepared 16, 21, 21, and 22 suspended graphene membranes with 0.01%, 1.1%, 50% and 99.2% $^{13}$C regions, respectively. Special care was given to select samples without grain boundaries, wrinkles or cracks. The measurement statistics at $T_m$~380 K is presented in Figure 3a. The solid lines in Figure 3a, fitted by the normal distribution, give about a 10% variance of *K*. The data scatter was mainly attributed to the variation in the optical absorption from graphene membrane to membrane and the uncertainty of measurement of *T* with the Raman spectrometer [10]. The average *K* at $T_m$≈380 K as a function of $^{13}$C concentration is plotted in Figure 3b. The 0.01% $^{13}$C graphene had the highest average *K* of 2805 W/m-K, which is 36% higher than natural abundance graphene (1.1% $^{13}$C) and 77% higher than 50% $^{13}$C. The room-temperature *K* values are correspondingly higher for all isotopic compositions. The maximum difference at room temperature is ~100% between *K* of 0.01% $^{13}$C and that of 50% $^{13}$C/$^{12}$C-mixture graphene.

Recently, MD simulations were used to compute the isotope effect on *K* of graphene [26]. The simulations predicted enhanced *K* for isotopically pure $^{12}$C or $^{13}$C graphene and strongly reduced *K* for mixed isotope graphene even with a small (~1%) addition of foreign isotopes. The MD models account well for contributions of the short-wavelength phonons but omit the long-wavelength phonons, which are important for heat transport in graphene [8-9, 14-16]. We modified a prior simulation approach by adding the long-wavelength phonon contributions from the Klemens' model [14-15]. The MD treatment of the short-wavelength phonons is essential for accurately describing the isotope scattering effects while Klemens' correction allows one to obtain *K* values for the large graphene samples. We used the optimized Brenner potential, also referred to as the reactive empirical bond order potential (REBO) [26]. It was shown from the solution of the Boltzmann transport equation (BTE) that the optimized Tersoff and Brenner potentials yield similar thermal conductivity values for graphene [27].

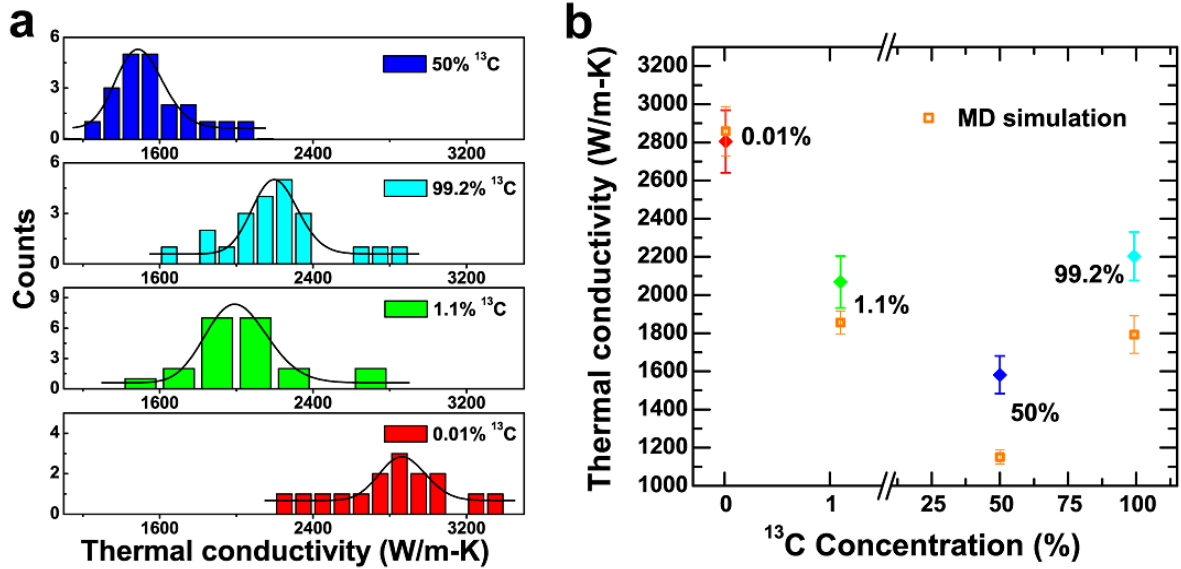

**Figure 3:** Thermal conductivity of graphene as a function of its isotopic composition. (a) Histogram of *K* of graphene films with 0.01%, 1.1%, 50% and 99.2% $^{13}$C isotopic concentration at ~380 K. The solid lines represent fits to the experimental data with the normal distribution. (b) The average value of the measured thermal conductivity *K* as a function of $^{13}$C concentration at ~380 K. The MD simulations results for *K* are shown with yellow squares for comparison.

At *T*~380 K, we obtained the following *K* from MD simulations (indicated in Figure 3b): 2859±129 W/mK, 1855±60 W/mK and 1151±38 W/mK for the isotopically pure $^{12}$C, 99% $^{12}$C (1% $^{13}$C) and for 50% $^{13}$C/$^{12}$C-mixture graphene, respectively. According to our MD simulations with the Klemens' correction for the long-wavelength phonons, *K* of the isotopically pure $^{12}$C graphene was ~40% enhanced as compared to the natural graphene, which is in line with our measurements. Some discrepancy may be related to wave interference effects characteristics for materials with large isotope impurity concentrations (>10%) [28]. The uncertainties of MD simulations (below 5%) indicated in Figure 3b were obtained from the Green-Kubo method, where *K* is computed by averaging the integral of the heat current autocorrelation function from ten uncorrelated micro-canonical (NVE) ensembles. The error bar characterizes the fluctuation of the averaged thermal conductivity value. It was also predicted theoretically that further reduction in *K* could be achieved if the isotopes were organized in small size clusters rather than being distributed randomly [29].

We have also compared our experimental results and MD simulations for the natural abundance (1.1% $^{13}$C) graphene with the predictions of the BTE models [8-9]. Plugging in *T*=380 K and a characteristic size *L*=2.8 μm to both numerical [8] and semi-analytical [9] models, we obtained *K*~2000 W/mK (with $\gamma_{LA}$=1.8 and $\gamma_{TA}$=0.75 as Gruneisen parameters for the longitudinal and transverse phonon modes in the semi-analytic model). This value is in quantitative agreement with the experimental and MD data points shown in Figure 3b.

It is interesting to note that the trend for *K* as a function of $^{13}$C concentration is in line with the predictions of the well-established virtual crystal model [30] used to calculate *K* in alloy semiconductors such as $Si_xGe_{1-x}$, $Al_xGa_{1-x}As$ or $Al_xGa_{1-x}N$. This model predicts the highest *K* for the material with either x~0 or (1-x)~0 and a fast decrease to a minimum as x deviates from 0. The fact that unlike in semiconductor compounds the phonon-point defect scattering in the isotopically modified graphene is limited to the mass-difference term (Eq. (1)) increases the value of the obtained experimental data for theory development.

**METHODS**

**Preparation of Isotopically Engineered Graphene**
Graphene films were grown on 25-μm thick Cu foils (Alfa Aesar) by a CVD method in a hot wall tube furnace at temperatures up to 1035$^o$C using a mixture of methane and hydrogen similar to the method reported previously. The carbon isotope-labeling [19] and the copper-foil enclosure [18] technique have been employed to synthesize the large-grain-size monolayer graphene with the regions of 0.01%, 1.1%, 50%, or 99.2% $^{13}$C on the same graphene sheet. The hydrogen flow rate was kept constant at 2 sccm with a partial pressure of 27 mTorr. The chamber background pressure was 17 mTorr. The methane flow rate was held at 1 sccm with the following dosing sequence: (1) 12 min for 99.99% $^{12}$CH$_4$ (Cambridge

Isotopes); (2) 15 min for 50%: 50% $^{12}CH_4/^{13}CH_4$ mixture; (3) 15 min for 98.9% $^{12}CH_4$ (Air Gas Inc.); (4) 15 min for 99.2% $^{13}CH_4$ (Cambridge Isotopes); (5) 3 min for 99.2% $^{13}CH_4$ (Cambridge Isotopes). The partial pressure from step 1 to 4 was about 36 mTorr, and a much higher partial pressure of 1 Torr was set for step 5 so as to fill the gaps between graphene domains to form a fully covered graphene layer on the surfaces on the inside of the enclosure[2]. The graphene-coated Cu foil enclosure was then carefully unfolded to expose what had been the inner surface (i.e., what had been the surface inside the 'pocket'), as this surface has particularly large grain size [18].

*Acknowledgements*

The authors appreciate comments by L. Shi, and H. Zhao. The work at UTA was supported by the National Science Foundation grant no.1006350 and the Office of Naval Research. The work at XMU was supported by the National Natural Science Foundation of China through grant nos. 91123009, 111104228, 10975115, 60827004 and 90921002 and the "973" program 2012CB619301 and 2011CB925600. The work at UCR was supported, in part, by the Semiconductor Research Corporation – Defense Advanced Research Project Agency FCRP Functional Engineered Nano Architectonic center, National Science Foundation and US Office of Naval Research. K.C. was supported by the NRF of Korea through WCU program grant no. R-31-2009-000-10083-0. RSR acknowledges support of W.M. Keck Foundation.

**Author Contributions** R.S.R. coordinated the project and data analysis; A.A.B. led the thermal data analysis; S.C. performed sample growth, measurements and data analysis; W.C. carried out the Raman optothermal measurement and data analysis. Q.W. assisted on the sample transfer. C.M. and J. K. contributed to the discussion of the data analysis. H.Z. and K.C. performed MD simulations; S.C., W.C., R.S.R. and A.A.B. wrote the manuscript.